\documentclass[aps,pre,superscriptaddress,groupedaddress,twocolumn]{revtex4}
\usepackage{graphicx}  
\usepackage{bm}        
\usepackage{amssymb}   
\usepackage{amsmath}
\usepackage{eufrak} 
\usepackage{amsthm}

\usepackage{enumerate}
\usepackage{color}
\usepackage{pgfplots}

\usepackage{accents}
\MakeRobust{\underaccent} 

\begin{document}

\title{ Comment on ``Thermodynamic uncertainty relation for time-delayed Langevin systems"}
\author{ M.L. Rosinberg}
\affiliation{Sorbonne Universit\'e, CNRS, Laboratoire de Physique Th\'eorique de la Mati\`ere Condens\'ee, LPTMC,\\  F-75005 Paris, France}
\email{mlr@lptmc.jussieu.fr}
\author{ G. Tarjus}
\affiliation{Sorbonne Universit\'e, CNRS, Laboratoire de Physique Th\'eorique de la Mati\`ere Condens\'ee, LPTMC,\\  F-75005 Paris, France}

\begin{abstract}
An extension of the thermodynamic uncertainty relation (TUR) to time-delayed Langevin systems has been recently proposed in \cite{VH2018}. Here we show that the derivation is erroneous.
\end{abstract} 


\maketitle

An important recent development in the field of stochastic thermodynamics has been the discovery of the so-called thermodynamic uncertainty relations (TURs) that provide general lower bounds on the fluctuations of time-integrated currents in nonequilibrium systems (see e.g. \cite{S2018} and references therein).  Such relations have been so far established for Markov processes  only but  in a recent work~\cite{VH2018} Vu and  Hasegawa  have presented an extension to time-delayed Langevin systems in a steady state. If correct, this would be an interesting result since delays are ubiquitous in real-world processes, for instance in biology. Unfortunately, the arguments  in \cite{VH2018} are incorrect, as we show in the present comment.

The  steady-state TUR for a  general Markovian dynamics is expressed as 
\begin{align}
\label{EqTUR}
\epsilon^2({\cal T})\equiv \frac{\langle \Theta^2\rangle-\langle\Theta \rangle^2}{\langle \Theta \rangle^2}\ge \frac{2}{\langle\Sigma \rangle}\ ,
\end{align}
where $\Theta$ is an arbitrary current integrated over some observation time ${\cal T}$ and $\langle\Sigma \rangle$ is the total entropy production accumulated by  ${\cal T}$ (in units where Boltzmann's constant is set to $k_B=1$). According to \cite{VH2018},  this relation remains valid for a time-delayed  Langevin dynamics provided $\langle\Sigma \rangle$ is replaced by a  ``generalized"  dissipation $\langle \Sigma_g\rangle$ (defined by Eq. (13) in \cite{VH2018} or Eq. (\ref{EqSigmag}) below).  This is an intriguing  result, but we here show that it follows from an incorrect treatment of  the  non-Markovian nature of the dynamics.
Specifically, the original Langevin equation for the $N$-dimensional random variable ${\bf x}(t)$ (cf. Eq. (2) in \cite{VH2018}), 
\begin{align}
\label{EqL}
\dot{\bf x}={\bf F}({\bf x},{\bf x}_{\tau})+\sqrt{2D}{\boldsymbol \xi}\ ,
\end{align} 
where $\tau$ is the delay, ${\bf x}_{\tau}\equiv {\bf x}(t-\tau)$, ${\bf F}({\bf x},{\bf x}_{\tau})$ is a drift force, and ${\boldsymbol \xi}$ is a Gaussian white noise,  has been mistakenly replaced by
\begin{align}
\label{EqLeff}
\dot{\bf x}={\overline {\bf F}}({\bf x})+\sqrt{2D}{\boldsymbol \xi}\ ,
\end{align} 
where ${\overline {\bf F}}({\bf x})$ is the (instantaneous) effective force defined by ${\overline {\bf F}}({\bf x})P^{ss}({\bf x})=\int {\bf F}({\bf x},{\bf x}_{\tau}) P^{ss}( {\bf x},t;{\bf x}_{\tau},t-\tau) \: d{\bf x}_{\tau}$
 (here, $P^{ss}({\bf x})$ and $P^{ss}( {\bf x},t;{\bf x}_{\tau},t-\tau)$ are the steady-state one time and two-time  probability distributions, respectively). This replacement allows Vu and  Hasegawa to express the probability density of a stochastic trajectory as ${\cal P}({\Gamma})\propto \exp \big[ -(1/4D)\int_{0}^{{\cal T}}\parallel \dot {\bf x}_t-{\overline {\bf F}}({\bf x}_t)\parallel^2dt\big]$ and  to obtain a lower  bound on $\epsilon^2({\cal T})$ by repeating the derivation performed in \cite{DS2018} for a Markovian Langevin dynamics. The  point we want to stress is that ${\cal P}({\Gamma})$ is {\it not} the probability of observing a  trajectory generated by the non-Markovian dynamics described by Eq. (\ref{EqL}) in a steady state, despite the fact that Eq.  (\ref{EqLeff}) leads to the same  probability distribution $P^{ss}({\bf x})$ as Eq. (\ref{EqL}). The same mistake was made in Ref. \cite{JXH2011} and signaled in \cite{RMT2015} where  ${\cal P}({\Gamma})$ as given above was shown to differ from the {\it exact} path probability computed for a linear time-delayed Langevin equation (in the case ${\cal T}\le \tau$).  In other words, we argue that the inequality derived in \cite{VH2018} applies to an effective  stochastic dynamics  that is not the true one. For the same reason,  and contrary to the  claim in Ref. \cite{JXH2011}, which is repeated in \cite{VH2018}, the quantity $\Delta S_g^{tot}$ (cf. Eq. (9) in \cite{VH2018}) does {\it not }satisfy an integral fluctuation theorem (IFT) with the actual dynamics described by (\ref{EqL}).  In fact, as shown in \cite{RMT2015}, there  is another candidate for the entropy production in time-delayed systems, which is obtained from time inversion and satisfies a proper IFT.
 
To illustrate our point, we  explicitly show that $2/\langle \Sigma_g\rangle$ is {\it not} a lower bound on the squared relative uncertainty $\epsilon^2({\cal T})$. To this aim, we consider a two-dimensional  version of Eq. (\ref{EqL}) with
\begin{align}
\label{EqForce}
{\bf F}({\bf x},{\bf x}_{\tau})=\left(
\begin{array}{c}
  -a_{11}x_1-a_{12}x_{2,\tau}\\
  -a_{21}x_{1,\tau}-a_{22}x_2
\end{array}
\right) \ , 
\end{align}
and we choose $\Theta=-\int_{0}^{{\cal T}} \{[a_{11}x_1(t)+a_{12}x_2(t)] \circ \dot x_1(t)+[a_{12}(t)x_1(t)+a_{22}x_2(t)]\circ \dot x_2(t)\}dt$ as the current, where  $\circ$  denotes the  Stratonovich product.  The model studied in section IV.C of \cite{VH2018} corresponds to the symmetric case $a_{11}=a_{22}$ and $a_{12}=-a_{21}$. We here focus on the model recently studied in \cite{RTM2018} in which there is no feedback from $1$ to $2$. Specifically, we take $a_{11}=a,a_{22}=b,a_{12}=-c$, and $a_{21}=0$. Note that these models are exactly solvable in a steady state due to the linearity of the force ${\bf F}({\bf x},{\bf x}_{\tau})$ and the Gaussian character of the white noise, which makes all probability distributions Gaussian. Therefore, there is no need to restrict the study to the small-$\tau$ limit, as done in \cite{VH2018}. 

In particular, using the same method as  \cite{FBF2003}, one can easily compute the steady-state correlation functions $\phi_{ij}(t)\equiv \langle x_i(0)x_j(t)\rangle$ for $0\le t\le \tau$. For instance, we find $\phi_{21}(t)=D c/[b(a+b)]e^{b(t-\tau)}$, from which we get
\begin{align}
\label{EqTheta}
\frac{1}{{\cal T}}\langle \Theta\rangle=c\dot \phi_{21}(0)=D \frac{c^2}{a+b}e^{-b\tau}\ .
\end{align}
The calculation of $\langle\Sigma_g \rangle$, defined in Ref. \cite{VH2018} as
\begin{align}
\label{EqSigmag}
\frac{1}{{\cal T}}\langle\Sigma_g \rangle\equiv \frac{1}{D}\langle {\overline {\bf F}}({\bf x}_t) \circ \dot {\bf x}_t\rangle\ ,
\end{align}
is also quite easy because the effective force is linear, i.e., ${\overline F}_1=-K_{11}x_1-K_{12}x_2, {\overline F}_2=F_2=-ax_2$, and the unknown coefficients $K_{11}$ and $K_{12}$ can be readily obtained by solving the steady-state Fokker-Planck  equation $\sum_{i=1,2}\partial_{x_i}[-{\overline F}_i({\bf x})P^{ss}({\bf x})+D\partial_{x_i}P^{ss}({\bf x})]=0$ where  $P^{ss}({\bf x})\propto \exp[-(1/2){\bf x}^T{\boldsymbol \sigma}^{-1}{\bf x}]$ and ${\boldsymbol \sigma}$ is the covariance matrix with elements $\sigma_{ij}\equiv \phi_{ij}(0)$. $K_{11}$ and $K_{12}$ are then expressed in terms of the $\sigma_{ij}$'s. This eventually yields $K_{11}=ab[(a+b)^2e^{b\tau}+c^2e^{-b\tau}]/[(ab+b^2+c^2)(a+b)e^{b\tau}-ac^2e^{-b\tau}]$, $K_{12}=-bc[(a+b)^2+c^2]/[(ab+b^2+c^2)(a+b)e^{b\tau}-ac^2e^{-b\tau}]$, and in turn 
\begin{align}
\langle\Sigma_g \rangle/{\cal T}&=\frac{bc^2[(a+b)^2+c^2]e^{-b\tau}}{(a+b)[(ab+b^2+c^2)(a+b)e^{b\tau}-ac^2e^{-b\tau}]}\ .
\end{align}
In the more general case of the force defined by Eq. (\ref{EqForce}), solving the Fokker-Planck equation does not fully determine $ {\overline {\bf F}}({\bf x})$, but one can then use the expression of the transition probability of Gaussian stationary processes in terms of the correlation functions (see Eq. (A1) in \cite{RTM2018}). (In passing, we also note that ${\overline {\bf F}}({\bf x})$  at the order $\tau$ is not obtained by simply taking the $\tau=0$ limit of the transition probability, as defined in Eq. (30)  in \cite{VH2018}. For instance, in the model considered in section IV.C of \cite{VH2018}, the exact calculation shows that the coefficient of $x_1$ in  ${\overline {\bf F}}_1$, and of  $x_2$ in ${\overline {\bf F}}_2$, is $-a+b^2\tau+{\cal O}(\tau^2)$. Accordingly, one should have $A=a-b^2\tau$ in the expression (43) of $P^{ss}({\bf x})$, implying that  the variance of $x_1$ and $x_2$ increases with $\tau$ instead of decreasing. This error suggests that the small-$\tau$ limit is also incorrect in the  two other examples considered in \cite{VH2018}. However,  this may be undetectable at the scale of the figures displayed in \cite{VH2018}.)

Finally, we compute the variance of $\Theta$, and for simplicity we focus on the long-time limit. Then $\lim_{{\cal T}\to \infty}{\cal T}^{-1}[\langle \Theta^2 \rangle-\langle \Theta\rangle^2]=\chi_{\Theta}''(0)$ where  $\chi_{\Theta}(k)$ is the scaled cumulant generating function  defined by $\chi_{\Theta}(k)=\lim_{{\cal T}\to \infty}{\cal T}^{-1}\ln \langle e^{k \Theta}\rangle $. A standard calculation using discrete Fourier series (see e.g. \cite{RH2016}) yields $\chi_{\Theta}(k)=-1/(2\pi)\int_0^{\infty} d\omega \ln [1- F_{k}(\omega)]$ with $F_{k}(\omega)= 4kD c^2\omega[a\sin(\omega \tau)+\omega\cos(\omega \tau)+k D\omega]/[(a^2+\omega^2)(b^2+\omega^2)]$. This  leads to 
\begin{align}
\chi_{\Theta}''(0)&=D^2\frac{c^2}{b(a+b)^3}\Big[(a+b)(2ab+2b^2+c^2)\nonumber\\
&+c^2[b(1+2b\tau)-a(1-2b\tau)]e^{-2b\tau}\Big]\ .
\end{align}
An example of the behavior of the quantity $R_{\Theta}\equiv \lim_{{\cal T}\to \infty}{\cal T}[\epsilon^2({\cal T})-2/\langle\Sigma_g \rangle]$ as a function of $\tau$ is shown in  Fig. 1. We observe that $R_{\Theta}$ becomes negative for large values of $\tau$, thus invalidating the TUR derived in \cite{VH2018} (more generally, the parabolic lower bound (25) on $\chi_{\Theta}(k)$ is invalid). On the other hand, as expected, $R_{\Theta}$ is always positive if $\epsilon^2({\cal T})$ is calculated with the effective stochastic dynamics defined by Eq. (3). We have confirmed these results by performing numerical simulations of the two dynamics.
\begin{figure}[hbt]
\begin{center}
\includegraphics[width=6.25cm]{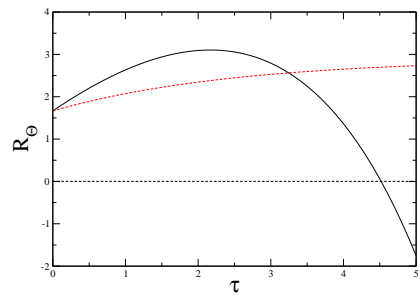}
 \caption{\label{Fig6} (Color on line) $R_{\Theta}\equiv \lim_{{\cal T}\to \infty}{\cal T}[\epsilon^2({\cal T})-2/\langle\Sigma_g \rangle]$ as a function of $\tau$ calculated with the original time-delayed dynamics [Eq. (\ref{EqL})] (solid black line) and the effective Markovian dynamics [Eq. (\ref{EqLeff})](dashed red line). The model parameters are $a=1,b=0.2, c=0.5$.} 
\end{center}
\end{figure}

In conclusion, the extension of the TUR to time-delayed Langevin systems is still an open problem. Whether or not the connection between TUR and Fisher information recently discussed in \cite{HV2018,D2018, ID2018} offers a possible solution remains to be seen.


\end{document}